\documentclass[twocolumn,prb,showpacs,amsmath,amstex,amssymb,mathfonts,superscriptaddress]{revtex4-1}

\usepackage{amsthm,color,amsfonts,graphicx,verbatim}
\usepackage{amsmath}
\usepackage{amssymb}
\usepackage{amsthm}
\usepackage{amsfonts}
\usepackage{listings}
\usepackage{enumerate}
\usepackage{latexsym}
\usepackage{bm}
\usepackage{graphicx}
\usepackage{dsfont}

\usepackage{hyperref}
\hypersetup{
    bookmarks=true,         
    unicode=false,          
    pdftoolbar=true,        
    pdfmenubar=true,        
    pdffitwindow=false,     
    pdfstartview={FitH},    
    pdftitle={Effective Hamiltonians, prethermalization and slow energy absorption in periodically driven many-body systems},    
    pdfauthor={},     
    pdfsubject={},   
    pdfcreator={},   
    pdfproducer={}, 
    pdfkeywords={keyword1} {key2} {key3}, 
    pdfnewwindow=true,      
    colorlinks=true,       
    linkcolor=black,          
    citecolor=blue,        
    filecolor=magenta,      
    urlcolor=blue           
}

\theoremstyle{plain}
\newtheorem{lemma}{Lemma}

\newcommand{\be}{\begin{equation}}
\newcommand{\ee}{\end{equation}}
\newcommand{\bea}{\begin{eqnarray}}
\newcommand{\eea}{\end{eqnarray}}

\newcommand{\la}{\langle}
\newcommand{\ra}{\rangle}

\renewcommand{\phi}{\varphi}
\renewcommand{\epsilon}{\varepsilon}
\newcommand{\str}{ |}
\newcommand{\norm}{ ||}

\begin{document}
\title{Effective Hamiltonians, prethermalization and slow energy absorption in periodically driven many-body systems}

\author{Dmitry A. Abanin}
\affiliation{{Department of Theoretical Physics, University of Geneva}}
\author{Wojciech De Roeck}
\affiliation{Instituut voor Theoretische Fysica, KU Leuven, Belgium}
\author{Wen Wei Ho}
\affiliation{{Department of Theoretical Physics, University of Geneva}}
\author{Fran\c{c}ois Huveneers}
\affiliation{CEREMADE, Universite Paris-Dauphine, France}


%
\date{\today}
\begin{abstract}

We establish some general dynamical properties of quantum many-body systems that are subject to a high-frequency periodic driving. We prove that such systems have a quasi-conserved extensive quantity $H_*$, which plays the role of an effective static Hamiltonian. The dynamics of the system (e.g., evolution of any local observable) is well-approximated by the evolution with the Hamiltonian $H_*$ up to time $\tau_*$, which is exponentially large in the driving frequency. We further show that the energy absorption rate is exponentially small in the driving frequency. In cases where $H_*$ is ergodic, the driven system prethermalizes to a thermal state described by $H_*$ at intermediate times $t\lesssim \tau_*$, eventually heating up to an infinite-temperature state after times $t\sim \tau_*$. Our results indicate that rapidly driven many-body systems generically exhibit prethermalization and very slow heating. We briefly discuss implications for experiments which realize topological states by periodic driving. 
 
\end{abstract}

\maketitle

%
%
%
%
%




\section{Introduction}
Recent advances in laser cooling techniques have resulted in experimental realizations of well-isolated, highly tunable quantum many-body systems of cold atoms~\cite{RevModPhys.80.885}. A rich experimental toolbox of available quantum optics, combined with the systems' slow intrinsic time scales, allow for a preparation of non-equilibrium many-body states and also a precise characterization of their quantum evolution. This has made the study of different dynamical regimes in many-body systems one of the forefront directions in modern condensed matter physics (for a review, see Ref.~\onlinecite{RevModPhys.83.863}). 

Conventional wisdom suggests that in a majority of many-body systems, the Hamiltonian time evolution starting from a non-equilibrium state should lead to thermalization at sufficiently long times: that is, physical observables reach thermal values, given by the microcanonical ensemble. Thermalization in such {\it ergodic} systems is understood in terms of the properties of individual eigenstates themselves -- observables measured in these eigenstates are already thermal, as  encapsulated by the eigenstate thermalization hypothesis (ETH) \cite{PhysRevA.43.2046, PhysRevE.50.888, 2008Natur.452..854R}. However, while ETH implies eventual thermalization, it does not make predictions regarding the intermediate-time dynamics of the system. Therefore, much work has been dedicated to studying how thermal equilibrium emerges in different many-body systems.

In particular, there is a class of systems, which exhibit the phenomenon of {\it prethermalization}~\cite{PhysRevLett.93.142002, PhysRevLett.103.056403, 1367-2630-12-5-055016, PhysRevA.81.033605}. Such systems have a set of approximate conservation laws, in addition to energy; therefore, at intermediate time scales they equilibrate to a state given by the generalized Gibbs ensemble, which is restricted by those conservation laws. Full thermal equilibrium is reached at much longer time scales, set by the relaxation times of the approximate integrals of motion. Prethermalization has been experimentally observed in a nearly integrable one-dimensional Bose gas~\cite{Gring1318}. 

In this paper, we establish some general properties of dynamics of periodically driven many-body systems (Floquet systems).  Periodic driving in quantum systems has recently attracted much theoretical and experimental attention, because, amongst many applications, it provides a tool for inducing effective magnetic fields, and for modifying topological properties of Bloch bands~\cite{PhysRevB.79.081406, PhysRevB.82.235114, 2011NatPh...7..490L, PhysRevX.3.031005}. Indeed, since periodic driving is naturally realized in cold atomic systems by applying electromagnetic fields, topologically non-trivial Bloch bands (Floquet topological insulators) in non-interacting systems have been observed experimentally\cite{PhysRevLett.111.185301,2014Natur.515..237J, 2015NatPh..11..162A}. However, since periodic driving breaks energy conservation, driven ergodic (many-body) systems are expected to heat up, eventually evolving into a featureless, infinite-temperature state~\cite{Ponte2015196, PhysRevX.4.041048, PhysRevE.90.012110, footnote1}. Thus, many-body effects are expected to generally make such Floquet systems unstable. Below, we derive general bounds for energy absorption rates in periodically driven many-body systems, which can be applied for instance to understand the lifetimes of Floquet topological insulators.

As the main result of the paper, we show that rapidly driven many-body systems with local interactions generally have a local, quasi-conserved extensive quantity, $H_*$, which plays the role of an effective Hamiltonian. At times $t\lesssim \tau_*$, the time evolution of any local observable is well-approximated by the Hamiltonian evolution with the time independent Hamiltonian $H_*$. Thus, assuming that the Hamiltonian $H_*$ is ergodic, the system exhibits prethermalization to a thermal state described by the Hamiltonian $H_*$, with an effective temperature set by the initial ``energy" $\la \psi_0| H_* |\psi_0\ra$. The quasi-conservation of $H_*$ is destroyed at timescale $t\sim \tau_*$, when energy absorption occurs and an infinite-temperature state is formed. We show that the heating timescale $\tau_*$ is exponentially large in the driving frequency $\omega$: 
\be\label{eq:tau}
\tau_* \sim  e^{c\frac{\omega}{h}},
\ee
where $c$ is a numerical constant of order 1, and $h$ has the meaning of a maximum energy per particle or spin,  precisely defined below. Thus, rapidly driven many-body systems generically have a very long prethermalization regime, and absorb energy exponentially slowly in the driving frequency. We emphasize that these results are non-perturbative; they generalize and complement our previous work, Ref.~\onlinecite{PhysRevLett.115.256803}, where bounds on linear-response heating rates were proven. As an implication of our result, we show that the measurement of a local operator time evolved with the effective Hamiltonian is close to the measurement of the same operator but exactly time evolved,  up to exponentially long times.

The structure of the rest of the paper is as follows. In Sec.~\ref{sec:setup}, we define the set-up and present the central idea of the transformation used to obtain our results. Then, in Sec.~\ref{sec:method}, we work out in detail, using the method presented, the optimal order $n_*$ of the transformation to obtain the effective Hamiltonian $H_*$ and also the heating time scale $\tau_*$ for which this effective Hamiltonian is valid. Next, in Sec.~\ref{sec:implications}, we present the implications of our result for the observation of a local operator. Lastly, we end with a discussion in Sec.~\ref{sec:conclusion}.

\section{Set-up and outline of method}
\label{sec:setup}
We consider a quantum many-body system subject to a drive with a period $T=2\pi/\omega$, described by a time dependent Hamiltonian:
\be\label{eq:Hamiltonian}
H(t)=H_0+V(t), \,\, V(t+T)=V(t),
\ee
where $H_0$ is time independent, and, without loss of generality, the time average of the driving term $V(t)$ is chosen to be zero, $\int_0^T V(t) \, dt=0$.  We focus on the case of a  lattice system with locally bounded Hilbert space. In other words, the Hilbert space of site $i$ is finite-dimensional, as is the case for fermions, spins, as well as hard-core bosons. We also restrict to one-dimensional systems, but this is not crucial to the method, see also Ref.~\onlinecite{2015arXiv150905386A}. Both $H_0$ and $V(t)$ are assumed to be local many-body operators, that is, they can be written as a sum of local terms: 
\be\label{eq:local}
H_0=\sum_{i} H_i, \,\,\, V(t)=\sum_i V_i(t), 
\ee
 where $i$ runs over all lattice sites, $i=1,...,N$. The locality of the interactions means that each term $H_i, V_i$ acts non-trivially on at most $R$ adjacent sites $i,i+1,\ldots, i+R-1$. (e.g., for the nearest-neighbor Heisenberg model, $R=2$); we refer to $R$ as the range of the operator. Each term $H_i, V_i$ is bounded by a constant interaction strength $h$:
 \be\label{eq:bounded_h}
 ||H_i ||\leq h, \,\, || |V_i(t)||\leq h. 
 \ee
We will focus on the case when the driving frequency is large (or equivalently, the driving period is small) compared to these local energy scales, that is, $hT \ll 1$.

Now, the unitary dynamics of the system is described by the time evolution operator $U(t)$, which obeys the equation:
\be\label{eq:U_equation}
i\partial_t U(t)=H(t) U(t), \,\, U(0)=I, 
\ee
where $I$ is the identity operator. 

Floquet theory (for a review, see Ref.~\onlinecite{2015AdPhy..64..139B}) predicts that the solution of Eq.(\ref{eq:U_equation}) can be written in the following form: 
\be\label{eq:FloquetH}
U(t)=P(t)e^{-iH_Ft}, 
\ee
where $P(t+T)=P(t)$ is a time periodic unitary such that $P(0)=I$, and $H_F$ is a time independent Floquet Hamiltonian. 
In particular, the evolution operator over one period is given by: 
\be\label{eq:evolution-one-period}
U(T)={\mathcal T} \exp\left(-i\int_0^T H(t)\, dt  \right)=e^{-iH_F T}. 
\ee 
Thus, the evolution of the system at stroboscopic times $t_n=nT, \, n\in \mathbb{Z}$ is governed by the time independent Hamiltonian $H_F$. Note that the choice of $H_F$ is not unique: given a particular $H_F$ and projectors $P_i=|i\ra\la i|$ onto its eigenstates with eigenvalues $E_i$, the Hamiltonian $H_F'=H_F+\sum _i  m_i \omega P_i$ is also a valid Floquet Hamiltonian for any $m_i \in \mathbb{Z}$.


Typically, there is no closed-form solution of Eq.(\ref{eq:FloquetH}), and one relies on iterative schemes such as the Magnus expansion to obtain $H_F$ for high-frequency drives (for a recent review, see Refs.~\onlinecite{2015AdPhy..64..139B, Blanes2009151, 1367-2630-17-9-093039}). In this approach, $H_F$ is expanded in terms of powers of $T$ (equivalently, of inverse frequency $1/\omega$), $H_F=\sum_n H_F^{(n)}$, where $H_F^{(n)}=O(T^n)$. The formal solution of Eqs.(\ref{eq:U_equation}, \ref{eq:FloquetH}) then gives $H_F^{(n)}$ expressed in terms of nested commutators of $H(t)$ at different times. However, the Magnus expansion is only known to converge for bounded Hamiltonians, such that $||H(t)||T\leq r_c$, with $r_c\sim 1$, $\forall t$~\cite{Blanes2009151}. 
Since many-body systems have extensive energies and do not satisfy this condition, the Magnus expansion is expected not to converge in this case. Indeed,  the existence of a quasi-local Floquet Hamiltonian $H_F$ would imply that the system does not heat up to an infinite-temperature state at long times, contrary to the general arguments based on the ETH~\cite{Ponte2015196}.


Therefore, we propose an alternative approach. The central idea is as follows: we unitarily transform the Hamiltonian, systematically removing time dependent terms at increasing order in $T$. Truncating the procedure at some optimal order $n_*$ (defined below), we obtain a quasi-conserved time independent Hamiltonian operator $H_*$. 

More concretely, we transform the system's wavefunction $|\psi(t)\ra$ by a time periodic unitary $Q(t+T)=Q(t)$, such that $Q(0)=I$:
\be\label{eq:Q}
|\phi(t)\ra=Q(t)|\psi(t)\ra. 
\ee
Importantly, the wavefunction $|\phi(t)\ra$ coincides with the original wavefunction $|\psi(t)\ra$ at stroboscopic times $t_n$. Its evolution is described by the Schroedinger equation
\be\label{eq:phi}
i\partial_t |\phi(t)\ra=H'(t)|\phi(t)\ra, 
\ee
with a modified Hamiltonian:
\be\label{eq:H'}
H'(t)=Q^\dagger H(t) Q -iQ^\dagger \partial_t Q. 
\ee
Thus, the transformation $Q(t)$ defines a new periodic Hamiltonian $H'(t)$, which gives the same stroboscopic evolution as the original Hamiltonian $H(t)$. 

For our purposes, it is convenient to write the operator $Q$ as an exponential of a periodic operator $\Omega(t)=\Omega (t+T)$, which is anti-Hermitian, $\Omega^\dagger=-\Omega$, and to represent $\Omega$ as an $n_{\rm max}$-degree polynomial in the driving period $T$:
\be\label{eq:Q_exp}
Q(t)=e^{\Omega}, \,\,\, \Omega=\sum_{q=1}^{n_{\rm max}}  \Omega_q, \,\, \Omega_q=O(T^q). 
\ee
Here, the order of the polynomial $n_{\rm max}$ should be treated as a parameter to be optimized in a manner described below. Using Duhamel's formula,  for $Q=e^{\Omega}$, Eq.~(\ref{eq:H'}) can be rewritten as follows: 
\be\label{eq:HOmega}
H'(t)=e^{-{\rm ad}_\Omega} (H_0+V(t))-i\frac{1-e^{-{\rm ad}_\Omega}}{{\rm ad}_\Omega} \partial_t \Omega, 
\ee
where ${\rm ad}_\Omega A=[\Omega,A]$, which gives an expansion of $H'(t)$ naturally in powers of $T$. 

We will show below that  the operators $\Omega_q$ can be chosen to get rid of the time dependence of $H'(t)$ of order $T^q$ for   $q \leq n_{\rm max}-1$, leaving behind a time-dependent piece of order $T^{n_{\rm max}}$. Furthermore, we will show that for a given $T$ and $H(t)$, there exists an optimal $n_{\rm max}=n_*$, for which this driving term's norm (suitably defined) becomes minimal.  For a many-body system with local interactions, we find that the optimal $n_* \sim  \omega$, and for this $n_*$, the driving term's norm is exponentially reduced by a factor of $e^{-c\frac{\omega}{h}}$. The time-independent part $H_*$ of the corresponding Hamiltonian $H'(t)$ then represents a quasi-conserved energy, valid for an exponentially long time $\tau_* \sim e^{c \frac{\omega}{h}}$.

\section{Method, optimal order, and heating time scale} 
\label{sec:method}
We now utilize  the transformation outlined in the previous section to transform the original Hamiltonian. We derive the optimal order $n_*$ at which the remaining driving term becomes minimal,  which gives us both the effective Hamiltonian $H_*$ and the heating time scale $\tau_*$.

\subsection{Simple example: single rotating frame transformation}
To get some familiarity regarding the use of our approach before going into full generality, it is instructive to first consider the simple example of a transformation $Q$ for $n_{\rm max}=1$, i.e., a single rotating frame transformation, so that $\Omega=\Omega_1=O(T)$, and $\Omega_1$ is chosen such that the driving term of order $T^0$ is eliminated in Eq.~(\ref{eq:HOmega}). 

Since the zeroth-order contribution in (\ref{eq:HOmega}) is given by $H_0+V(t)-i\partial_t\Omega_1$, we define $\Omega_1$ by:
\be\label{eq:Omega1}
\Omega_1(t)=-i\int_0^t V(t')\, dt'.
\ee
With this choice of $\Omega_1$,  $H'$ of Eq.~(\ref{eq:HOmega}) can be expanded in ``powers of $T$'', 
\be\label{eq:H'1}
H'(t)=\sum_{q=0}^\infty H^{(q)}(t),
\ee
where $H^{(q)}(t)$ is the term of order $T^q$:
\be\label{eq:H'higher}
H^{(q)}(t)= \frac{(-{\rm ad}_{\Omega_1})^q}{q!} H_0+\frac{q(-{\rm ad}_{\Omega_1})^q}{(q+1)!} V(t). 
\ee
To the first order in $T$, the rotated Hamiltonian $H'$ is given by: 
 $$
 H'(t) =  H_0+\bar{H}^{(1)}+V^{(1)}(t)+O(T^2), 
 $$
 where $\bar{H}^{(q)}=\frac{1}{T}\int_0^T H^{(q)}(t) dt$ is the time-independent part of $H^{(q)}(t)$, and $V^{(q)}(t)=H^{(q)}(t)-\bar{H}^{(q)}$ is the new driving term (with zero time-average) at this order. A straightforward calculation shows that $\bar{H}^{(1)}=\frac{T}{2}\int_0^T dt_1 \int_0^{t_1} dt_2 [H(t_1),H(t_2)]$ coincides with the second order of the Magnus expansion. 

We see that $H$ is the order $T^0$ piece of $H'(t)$ and is time independent, while the remaining piece $\delta H'(t) \equiv \sum_{q \geq 1} H^{(q)}(t)$ that appears at orders $T^1$ and higher is still time dependent, and represents the new driving term. Thus the rotated Hamiltonian can be written as
\be
H'(t) = H_0 + \delta H'(t), \qquad \delta H'(t) = O(T^1).
\ee
Contrasted to the original Hamiltonian $H + V(t)$, it appears that the new driving term's norm has been reduced by a factor of $T$. 

However, there is an important distinction to be made between $H'(t)$ and the original Hamiltonian $H+V(t)$. In a many-body system, the rotated Hamiltonian $H'$ in Eq.(\ref{eq:H'1}) is now quasi-local instead of being strictly local. This is because $H^{(q)}(t)$ involves $q$ nested commutators of $\Omega_1$ and $H,V(t)$, and the norm of each term decreases exponentially with $q$ for sufficiently rapid driving. To establish this, we note that each term $H^{(q)}(t)$ is extensive and can be written as $H^{(q)}(t)=\sum _i H^{(q)}_i (t)$. We denote the maximum local(l) norm of $H_i^{(q)}(t)$ as $|| H^{(q)}(t)||_l\equiv \sup_i ||H^{(q)}_i (t) ||$, and use the following fact: for any two extensive operators $A=\sum _i A_i$, $B=\sum _i B_i$ of range $R_A, R_B$, respectively, such that $||A ||_l \leq a$, $||B ||_l \leq b$, $C={\rm ad}_A B$ has a range of at most $R_C=R_A+R_B-1$, and $C=\sum _i C_i$, with norm
  \be\label{eq:ad_estimate}
  ||C||_l\leq  2(R_A+R_B-1)ab.
 \ee 
 This is because each operator $A_i$ can commute non-trivially with at most $R_A+R_B-1$ operators $B_j$. Repeatedly applying this estimate to the operators ${\rm ad}_{\Omega_1}^q H$, ${\rm ad}_{\Omega_1}^q V(t)$ that enter Eq.(\ref{eq:H'higher}), and using the fact that $\Omega_1$ has range $R$, and $||\Omega_1||\leq hT$ (which follows from Eq.(\ref{eq:Omega1})), we obtain: 
 \be\label{eq:Hk_estimate}
 ||H^{(k)}(q) ||_l \leq 2h (2hRT)^q, 
 \ee
and the range of $H^{(q)}(t)$ equals $q(R-1)+R$. Thus, this  establishes the quasi-locality of the Hamiltonian $H'(t)$. 

Further, by using an appropriately weighted local norm (see appendix (\ref{appendix:bound}) and also Ref.~\onlinecite{2015arXiv150905386A}), the size of $\delta H'(t)$ is $O(T)$. Therefore, we see that the transformation $\Omega_1$ reduces the amplitude of the time dependent term by a factor of order $T$, while at the same time making the Hamiltonian quasi-local, and renormalizing its time independent part. 

\subsection{General case}
Next, we proceed to the general case of $n_{\rm max}>1$. Then, $\Omega=\sum_{p=1}^{n_{\rm max}} \Omega_q$, where as mentioned, $\Omega_q=O(T^q)$ is chosen such that the only time dependent terms in the Hamiltonian $H'$ are of order $T^{n_{\rm max}}$.  This condition gives us a set of recursive relations for $\Omega_{q}(t)$: we use them to `absorb' the time dependent pieces of order $T^q$ in $H'$ for $1\leq q \leq n_{\rm max}-1$. To derive these relations, we first note that the term of the order $T^q$ in $H'(t)$ has the following form: 
\begin{widetext}
\be\label{eq:term_q}
H^{(q)}(t)=G^{(q)}(t)-i\partial_t \Omega_{q+1}(t),
\ee
where $G^{(q)}(t)$ is expressed in terms of $\Omega_1(t),...,\Omega_q(t)$: 
\be\label{eq:G_q}
G^{q}(t)=\sum_{k=1}^{q} \frac{(-1)^k}{k!} \sum_{\begin{subarray}{l}{1\leq i_1,...,i_k\leq q} \\ { i_1+...+i_k=q} \end{subarray}} 
{\rm ad}_{\Omega_{i_1}}...{\rm ad}_{\Omega_{i_k}} \, H(t) +
 i\sum _{m=1}^q \sum_{k=1}^{q+1-m} \frac{(-1)^{k+1}}{(k+1)!} \sum_{\begin{subarray}{l}{1\leq i_1,...,i_k\leq q+1-m} \\ { i_1+...+i_k=q+1-m} \end{subarray}} 
{\rm ad}_{\Omega_{i_1}}...{\rm ad}_{\Omega_{i_k}} \, \partial_t \Omega_m(t). 
\ee
\end{widetext}
(For $q \geq n_{\rm max}$, we set $\Omega_{q>n_{\rm max}}\equiv 0$.) We can separate $G^{(q)}(t)$ into a time independent part, $\bar{H}^{(q)}$, and a time dependent part $V^{(q)}(t)$ with zero average over one period: 
\be\label{eq:Hq}
\bar{H}^{(q)}=\frac 1T \int_0^T G^{(q)}(t)\, dt, \,\, V^{(q)}(t)=G^{(q)}(t)-\bar{H}^{(q)}. 
\ee
We eliminate the time dependent term of the order $T^q$ in $H'(t)$ (see Eq.(\ref{eq:term_q})) by choosing $\Omega_{q+1}(t)$ as follows:
\be\label{eq:Omega_q}
\Omega_{q+1}(t)=-i\int_0^{t} V^{(q)}(t') \, dt'
\ee
for $q\leq n_{\rm max}-1$. In particular, for $q=0$, $\Omega_1(t)$ is given by Eq.(\ref{eq:Omega1}). 

Relations (\ref{eq:G_q},\ref{eq:Hq},\ref{eq:Omega_q}) define the transformation $\Omega$ which makes the time dependent terms in the Hamiltonian $H'$ of the order $T^{n_{\rm max}}$:
\be\label{eq:H'simple}
H'(t)=H_0+\sum_{q=1}^{\rm n_{\rm max}-1} \bar{H}^{(q)} +\delta H' (t), \,\, \delta H'(t) = O(T^{n_{\rm max}}). 
\ee
In a manner similar to the simple example of $n_{\rm max}=1$ considered before, the full Hamiltonian $H'(t)$ and time dependent term $\delta H' (t)$ can be shown to be quasi-local (see appendix (\ref{appendix:bound})). 

Now, let us now estimate the norm of $\delta H'(t)$. We argue that there is an optimal $n_{\rm max}$ which we call $n_*$, for which the procedure we have outlined before approximatively minimizes the local norm of $\delta H'(t)$. This has physical consequences for both the heating time scale and the observation of a local observable, for example. Thus, $n_{\rm max}$ should be chosen as $n_*$. 


To this end, we prove a number of inequalities for the norms of various operators, $||G^{(q)}||_l, ||\Omega_q||_l, ||\bar{H}^{(q)}||_l$ and $|| V^{(q)}||_l$ (refer to the appendix and to Ref.~\onlinecite{2015arXiv150905386A} for generalizations). In the following, there will appear constants $C, c$, etc., which depend on the microscopic details of the system such as $h$ and $R$, but importantly not on the driving period $T$. It is to be understood that these constants can be different for different objects in question that are being bounded. Now, for $q \leq n_{\rm max}-1$, we have
 \be\label{eq:G_growth}
||G^{(q)}||_l \leq (C_0R)^q q! h (hT)^q,
\ee
with $C_0$ a combinatorial constant of order $1$. 
The other operators have then derived bounds since $||\bar{H}^{(q)}||_l \leq ||G^{(q)}||_l$ and $  || V^{(q)}||_l \leq 2||G^{(q)}||_l $. For $\Omega_q$, we have
\be\label{eq:H_estimate}
 ||\Omega_{q+1}||_l \leq 2(C_0R)^q q!  (hT)^{q+1}.
\ee
The $q!$ factor in the above bounds arises because of the many-body nature of the system: $G^{(q)}$ involves $q$ nested commutators of $H_0$, $V(t)$. 
Eq.~(\ref{eq:G_growth}) shows that there are two competing effects which control the behavior of $||G^{(q)}||_l$: suppression of $||G^{(q)}||_l$ by a factor of $T^q$, and its growth due to $q!$. Eventually, the factorial dominates and therefore for $q> \frac{1}{C_0RhT}$, the local norm of  $G^{(q)}$ stops decreasing with $q$.

The optimal $n_{\rm max}$  that we have to choose for is roughly the same as the one to choose to minimize the norm of $\Omega_{n_{\rm max}}$ or $G^{(n_{\rm max})}$ (see appendix). From the right hand side of Eq.~(\ref{eq:G_growth}) or Eq.~(\ref{eq:H_estimate}), we obtain
\be\label{eq:n_optimal}
n_* = \frac{e^{-r}}{C_0e RhT},
\ee
with $r=r(R)$ (independent of $T$) defined in appendix (\ref{appendix:bound}).   This gives us  the following bound on $\Omega_q$,
\be
|| \Omega_{q} ||_l \leq C e^{-r {q} },
\ee
for $q \leq n_*$, which in turn gives us an estimate on the remainder:
\be\label{eq:bound remainder}
\norm \delta H'(t)  \norm_l \leq C e^{-cn_*}.
\ee 
As already indicated, the truly useful version of this bound also expresses that local terms in $\delta H(t)$ with large range are additionally damped, and this is indeed captured by the use of a  stronger norm in the appendix. 

Furthermore, at this optimal order, 
 the time independent part of the transformed Hamiltonian is a physical, local many-body Hamiltonian
\be
H_* \equiv H_0+\sum_{q=1}^{\rm n_{\rm max}-1} \bar{H}^{(q)},
\ee
and differs from the original Hamiltonian $H$ by
 a sum of small local terms, more precisely
\be 
\label{eqn:HeffSmall}
\tfrac{1}{N} ||  H_*-H_0   || \leq  \sum_{q=1}^{n_{*}-1}  ||  \bar{H}^{(q)} ||_{l}    \leq Ch.
\ee
. 

Eq.~\eqref{eq:bound remainder} together with \eqref{eqn:HeffSmall} imply that the energy absorption rate (per volume) is exponentially small, giving us a characteristic heating time scale that scales like
\begin{align}
\tau_* \sim e^{c \frac{\omega}{h}}.
\end{align}
The operator $H_*$ is therefore a quasi-conserved extensive quantity, playing the role of an effective static Hamiltonian,  and it can be used to accurately describe stroboscopic dynamics up to times $\tau_*$.

\section{Implications: evolution of a local observable}
\label{sec:implications}
Next, we spell out the consequences of the existence of this effective Hamiltonian $H_*$ for the time evolution of a local observable $O$, with $ || O  ||=1$. Let us consider the difference between $O$ evolved in time using the exact time evolution operator and the time evolution generated by the effective Hamiltonian $H_*$. The difference 
\be \label{eq: difference}
Q(t)U^\dagger(t)OU(t)Q^\dagger(t) -   e^{i t H_*}O e^{-i t H_*}
\ee
can be recast, using the frame transformation and Duhamel's formula, as
\be  \label{eq: duhamel bound}
  i \int_0^t d s   \, W^*(s,t)      [\delta H'(s),    e^{i s H_*}O e^{-i s H_*}  ] W(s,t),
\ee
where $W(s,t)=W^{-1}(s) W(t)$ is the evolution from time $s$ to $t$ generated by the time dependent Hamiltonian $H'(t)$.  The norm of the difference can be bounded using the unitarity of $W(s,t)$ as 
\be  \label{eq: lr bound norm}
\int_0^t d s ||  [\delta H'(s),    e^{i s H_*}O e^{-i s H_*}  ]  || 
\ee
which can be controlled by the Lieb-Robinson bound, see Refs.~\onlinecite{Lieb1972, 2010arXiv1004.2086N}. Indeed, let us first pretend that the range of local terms in $\delta H'(s)$ is maximally $Rn_*$, then the Lieb-Robinson bound yields
\be  \label{eq: lr bound}
||  [\delta H'(s),    e^{i s H_*}O e^{-i s H_*}  ]  ||   \leq  C|| \delta H'(s)||_l( s v_* + Rn_*)
\ee
where $v_*$ is the Lieb-Robinson velocity of $H_*$, which can be chosen to be $\sim Ch$. Here $C$ is a numerical constant of order $1$. 

The bound \eqref{eq: lr bound} expresses that only those terms in $\delta H'(s)$ that have support within distance $sv_*$ of the support of $O$, contribute to the commutator, see Ref.~\onlinecite{2015arXiv150905386A} for a more detailed derivation of such bounds. Since in our case the support of local terms in $\delta H'(s)$ can grow arbitrarily large (because it is quasi-local), we however need to use the exponential decay in support of the norm of each local term in $\delta H'(t)$ to derive Eq.~\eqref{eq: lr bound}, in which case $C$ depends on the decay constant. We omit this straightforward calculation and refer to Ref.~\onlinecite{2015arXiv150905386A}.

Using $|| \delta H'(s)||_l \leq Ce^{-cn_*}$, we conclude that the difference Eq.~\eqref{eq: difference} grows as  $\sim t^2 e^{-cn_*}$ with $t$ and hence it remains  small up to an exponentially long time $t \sim   e^{ c\frac{ \omega}{h}}$. Thus, a measurement of $O(t)$ that is time evolved by the effective Hamiltonian $H_*$ will be close to the measurement of $O(t)$ that is time evolved by the exact Hamiltonian $H + V(t)$, for an exponentially long time.

%
%

\section{Discussion and conclusion}
\label{sec:conclusion}
In this paper, we considered many-body systems subject to a high-frequency periodic driving. We have shown that there is a broad time window, $t\lesssim \tau_*$, in which stroboscopic dynamics of such systems is controlled by an effective time independent Hamiltonian $H_*$. We have used a series of ``gauge", time periodic unitary transformations to effectively reduce the strength of the driving term and to establish the existence of $H_*$. The advantage of our approach compared to the standard Magnus expansion~\cite{2015AdPhy..64..139B, Blanes2009151, 1367-2630-17-9-093039} is that it allows us to control the magnitude of the driving terms after the transformations. 

We note that recently Canovi et al.~\cite{PhysRevE.93.012130} and Bukov et al.~\cite{PhysRevLett.115.205301} discussed prethermalization in {\it weakly interacting} driven systems. Our results complement these works: we have shown that (rapidly) driven interacting systems {\it generically} exhibit a broad prethermalization regime, which can be observed in a quench experiment as follows. Let us initially prepare the system in some non-equilibrium state $|\psi\rangle$, and subject it to a rapid periodic drive. At times $t\lesssim \tau_*$ the system will reach a steady state, in which physical observables have thermal values, $\langle \psi(t)| O|\psi(t)\rangle={\rm Tr} \left( O \rho \right)$, where the density matrix $\rho\propto e^{-H_*/T_{\rm eff}}$, with $T_{\rm eff}$ being the effective temperature set by the energy density of the initial state. Thus, at times $t\lesssim \tau_*$ the system appears as if it is not heating up. The system will absorb energy and relax to a featureless, infinite-temperature state beyond times $t\sim \tau_*$. We expect this phenomenon to be observable in driven system of cold atoms and spins (assuming relaxation of spins due to phonons is slow).  


Finally, we briefly discuss the implications of our results for the current efforts to realize topologically non-trivial strongly correlated states (e.g., fractional Chern insulators) in periodically driven systems. Experimentally, one tries to design a drive for which the ground state of an effective time independent Hamiltonian (usually calculated within low-order Magnus expansion) is topologically non-trivial. A central challenge is to prepare the system in a ground state of the effective Hamiltonian. Since we have shown that the dynamics of the system is controlled by $H_*$ up to exponentially long times, one can envision that the ``Floquet fractional Chern insulators" can be prepared as follows. Let us assume that the system can be initially prepared in a (topologically trivial) ground state of the Hamiltonian $H$. Then, the driving is switched on adiabatically to the value which corresponds to the desired effective Hamiltonian $H_*$. However, the switching should also be done quickly compared to $\tau_*$ to avoid energy absorption. Since $H$ and $H_*$ describe different phases, the system will necessarily go through a quantum critical point (QCP), and excitations will be created via a Kibble-Zurek mechanism. The  number of excitations can be minimized by designing a non-linear passage through the QCP~\cite{PhysRevLett.101.076801}. We leave a detailed exploration of these ideas for future work\cite{2016arXiv161105024H}.

{\it Note added.} Recently, a related result, Ref.~\onlinecite{PhysRevLett.116.120401} appeared, building on Ref.~\onlinecite{Kuwahara201696} (local driving).  Ref.~\onlinecite{PhysRevLett.116.120401} proves a similar bound for the absorption rate in driven systems using a different approach (namely, studying evolution over one driving period).

\section{Acknowledgements}
D.A.\@ acknowledges support by Alfred Sloan Foundation.  
W.D.R.\@ also thanks the DFG (German Research Fund), the Belgian Interuniversity Attraction Pole (P07/18 Dygest), as well as ANR grant JCJC for financial support.

\bibliography{refs}

\begin{thebibliography}{37}
\bibitem [{foo()}]{footnote1}%
  \BibitemOpen
  \href@noop {} {}\bibinfo {note} {It should be noted that driven many-body
  localized systems can exhibit dynamical localization in the energy space,
  avoiding heating to an infinite-temperature state, see
  Refs.~\onlinecite{Ponte2015196, PhysRevLett.115.030402,
  PhysRevLett.114.140401, Abanin20161}}\BibitemShut {NoStop}%
\end{thebibliography}

\setcounter{equation}{0}
\renewcommand{\theequation}{S\arabic{equation}}
\renewcommand{\thefigure}{S\arabic{figure}}

\begin{widetext}

\appendix
\section{Technical estimates and proofs}

Here, we provide a proof of the bounds on the terms of the renormalized Hamiltonian $H'$ and the bound on the remainder $\delta H(t)$.

\subsection{Setup}
Let us first recall the norm that we are using. We write operators $B=B(t)$, periodic in time, as a sum of local terms
$$
B(t)=\sum_i B_i(t)
$$
where $i$ runs over the sites of the (finite but large) volume and $B_i$ is an operator that acts nontrivially on the sites $i,i+1, \ldots, j$ where $j-i < R_B $, with $R_B$ independent of $i$ and called the ``range of $B$''. The local norm $ || B ||_l $ is then defined as 
 $$ 
 || B ||_l = \sup_i \sup_t || B_i(t) ||.
 $$
 In what follows, we mostly drop the dependence on $t$ from the notation. 
Let us now list the important bounds on local norms that we claim:    The operators $V^{(q)},G^{(q)}, \bar{H}^{(q)}, \Omega_{q+1}$ have range $R(q+1)$ and their local norms are bounded as
 \begin{equation}\label{eq:G_growthapp}
||G^{(q)}||_l \leq  q! h (C_0RhT)^q,
\end{equation}
\begin{equation}\label{eq:H_estimateapp}
|| {\bar H}^{(q)}||_l \leq   ||G^{(q)}||_l, \qquad    || V^{(q)}||_l \leq 2 ||G^{(q)}||_l. 
\end{equation}
\begin{equation}\label{eq:omega_estimateapp}
 ||\Omega_{q+1}||_l \leq T  || V^{(q)}||_l  \leq 2(C_0R)^q q!  (hT)^{q+1}.
\end{equation}
 We write $G^0\equiv H$, so that \eqref{eq:G_growthapp} is consistent with the fact that the range of the original operator $H$ is $R$ and its local norm is $h$. 
The bounds in  \eqref{eq:H_estimateapp} follow immediately because time averaging of local term does not increase its norm (here we use that the norm was defined as the supremum over time).  The bound \eqref{eq:omega_estimateapp} follows from \eqref{eq:G_growthapp} for a given $q$ because the integral over one period yields an additional factor $T$, i.e. using $ ||  \int_0^T d t B_i(t)  ||  \leq T\sup_t ||  B_i(t) ||   $. 
 Hence we have now in particular established the above bounds for $q=0$ and we have shown that the bound \eqref{eq:G_growthapp}  implies the others, for a given $q$. Therefore, to complete an inductive proof, it suffices to prove \eqref{eq:G_growthapp}  for $q$ while assuming the other bounds for  all $ q' <q$.  To achieve this, we use  \eqref{eq:G_q}: 
\begin{equation}\label{eq:G_qapp}
G^{(q)} (t)=\sum_{k=1}^{q} \frac{(-1)^k}{k!} \sum_{\begin{subarray}{l}{1\leq i_1,...,i_k\leq q} \\ { i_1+...+i_k=q} \end{subarray}} 
{\rm ad}_{\Omega_{i_1}}...{\rm ad}_{\Omega_{i_k}} \, G^{(0)} +
 i\sum _{m=1}^q \sum_{k=1}^{q+1-m} \frac{(-1)^{k+1}}{(k+1)!} \sum_{\begin{subarray}{l}{1\leq i_1,...,i_k\leq q+1-m} \\ { i_1+...+i_k=q+1-m} \end{subarray}} 
{\rm ad}_{\Omega_{i_1}}...{\rm ad}_{\Omega_{i_k}} \, V^{(m-1)}. 
\end{equation}
The right hand side of \eqref{eq:G_qapp} is a sum of local operators, all of which have support not greater than $R(q+1)$ adjacent sites.  
We estimate the norm of each of these local operators by using repeatedly $\norm [A_i,B_j]\norm \leq 2 \norm A_i \norm \norm B_j \norm$ (if $A_i,B_j$ have overlapping support) and the  above bounds for $ q' <q$. 
Then we sum the bounds on all terms that have site $1$ as the leftmost site of their support, to get a bound for $\norm G^{(q)}\norm_l$.  The result is,  separately for the first (\ref{eq:commone}) and second (\ref{eq:commtwo}) term of \eqref{eq:G_qapp}: 
\begin{equation}\label{eq:commone}
K(q)\sum_{n= 2}^{q+1} \frac{4^{n}(C_0R)^{1-n}}{(n-1)!}  \,   \sum^{(R,q)}_{\{I_j\} } \chi\left(\frac{\str I_1\str}{R} =1 \right)   \prod_{j=1}^n  \left(\frac{\str I_j\str}{R}-1 \right)! 
\end{equation}
\begin{equation}\label{eq:commtwo}
K(q)  \sum_{n= 2}^{q+1} \frac{4^n(C_0R)^{1-n}}{n!}   \,  \sum^{(R,q)}_{\{I_j\} }   \prod_{j=1}^n  \left(\frac{\str I_j\str}{R}-1 \right)! 
\end{equation}
where $\chi(A)=1$ if statement $A$ holds true, and $0$ otherwise, and we abbreviated 
$$
K(q)= h(C_0RhT)^q.
$$
The sum $  \sum^{(R,q)}_{\{I_j\}}$ is over all sequences   $I_j, j=1,\ldots, n$ of discrete intervals (sets of adjacent sites) $ I_j \subset  \mathbb{N}$ such that we have the following conditions.
\begin{enumerate}
\item All interval lengths $\str I_j \str$ are multiples of $R$: $\str I_j \str  \in R \mathbb{N}$.
\item  For $j>1$,  $I_j \cap \left(\cup_{i=1}^{j-1}  I_i \right)$ is nonempty.
\item $\sum_j \str I_j \str =R(q+1)$.
\item  $\min (\cup_{i=1}^{n}  I_i )=1$
\end{enumerate}
Intersection condition $2$ stems from the structure of nested commutators. Condition 4 says that we consider terms the support of which starts at site $1$.

To conclude the proof of the bounds \eqref{eq:G_growthapp}, we have to show that, for some ($q$-independent) choice of the constant $C_0$, the sum of \eqref{eq:commone} and \eqref{eq:commtwo}  is bounded by $K(q) q!$. 
It is sufficient to prove a bound on \eqref{eq:commtwo}, as \eqref{eq:commone} reduces to that case upon increasing $C_0\to C_1 C_0$, with $C_1$ such that $\frac{C_1^{1-n} }{(n-1)!} \leq 1/n!$. For the same reason, we can replace $C_0^{1-n}$ by $C_0^{-n}$. 
Hence, we show Lemma 1.
\begin{lemma} \label{lem: main} For some $C_0$ independent of $q,R$, 
\be  \label{eq: basiclem}
\frac{1}{q!}\sum_{n=2}^{q+1} \frac{(C_0R)^{-n}}{n!}    \sum^{(R,q)}_{\{I_j\} }    \prod_j \left(\frac{\str I_j\str}{R}-1 \right)!    \leq 1.
\ee
\end{lemma}
\subsection{Proof of Lemma \ref{lem: main}}

For simplicity, we set $R=1$. The case $R>1$ follows analogously. 
We set $m_j: =\str I_j \str$ and we let $L$ be such that $\cup_{i=1}^{n}  I_i =[1,L]$. Note that $1 \leq L \leq q$ because there are at least two overlapping intervals and the sum of their lengths is $q+1$. 
We write $m=(m_j)_{j=1}^n \in [1,L]^n$ for the sequence of lengths. 
First, we dominate the sum on the left-hand side of \eqref{eq: basiclem} as
\be  \label{eq: m here}
 \sum_{n=2 }^{q+1}   \sum_{L = 1}^q \frac{nC_0^{-n} }{n!q!}   L^{n-1} \sum_{m:\, \sum_{j}m_j=q+1}   \prod_j   (m_j-1)!
\ee
where the factor $L^{n-1}$ accounts for the choice of  position of the intervals in the stretch $[1,L]$.  There are $n$ intervals but at least one of them has to be placed such that its leftmost point is at $1$, therefore we have $nL^{n-1}$ instead of $L^n$. 

We use the upper bound in Stirling's formula
$$
c (N/e)^{N+1/2}  \leq N!  \leq C (N/e)^{N+1/2}
$$
to get 
\be \label{eq: first stirling}
(m_j-1)! \leq  C (m_j/e)^{m_j -1/2}.
\ee

 Here and below we use $c,C$ for numerical constants that do not depend on $q$, their value can change from line to line.  To deal with the product over such factors we define $Z(L,n)= \sum_{n_{\cal L} =0}^{n}  Z(L,n,n_{\cal L})    $ with
$$
Z(L,n,n_{\cal L})  :=   \sum_{m: \, \sum m_j=q+1}  \,   \chi(n_{\cal L}(m) =n_{\cal L})    \prod_j (m_j/L)^{m_j-1/2} 
$$
where $0 \leq n_{\cal L}(m) \leq n$ is the number of 'large' naturals in the sequence $m$: 
$$
n_{\cal L}(m) :=  \str \{ j \in \{1,\ldots, n\} \, \str \,  m_j \geq \alpha L    \}\str, \qquad \text{for some fixed $\alpha$ with $1/2 <\alpha < 1$}.
$$
Plugging \eqref{eq: first stirling} into \eqref{eq: m here} and using the above notation, we get
\begin{align}
\text{\eqref{eq: m here}}  \, & \leq   \, \sum_{n=2 }^{q+1} \sum_{L = 1}^q  \frac{C^nC_0^{-n}}{n!q!}   L^{n-1}      \frac{L^{q+1-n/2}}{e^{q} }  Z(L,n)  \\
 & \leq   \,    \sum_{n=2 }^{q+1}  \sum_{L = 1}^q \frac{(C/C_0)^{n}}{n!}   \frac{L^{n/2} }{\sqrt{q}} (L/q)^{q}  \,  \sum_{n_{\cal L} =0}^{n}  Z(L,n,n_{\cal L}),  \label{eq: first zet}
\end{align}
where we have also used the lower bound in Stirling's formula. For $Z(L,n,n_{\cal L})$, we find the  bound
\be  \label{eq: conclusion zet}
Z(L,n,n_{\cal L})  \leq     \frac{C^nn!}{n_{\cal S} ! n_{\cal L} ! }  \,   \chi (  q \geq \alpha n_{\cal L} L +n_{\cal S} -1) \,   (1/L)^{n_{\cal S}/2}, \qquad   n_{\cal S}\equiv n-n_{\cal L}
\ee
In words, short intervals yield small factors $1/L$, but $n_{\cal L}$ constrains $q,L$.

\emph{Proof of \eqref{eq: conclusion zet}.} 
Note that $\sum_j m_j=q+1$ implies $q \geq \alpha n_{\cal L} L +n_{\cal S} -1$. 
There are  $\frac{n!}{n_{\cal S} ! n_{\cal L} ! }$  ways to choose $n_{\cal L}$ large naturals from $n$. This yields the bound
$$
Z(L,n,n_{\cal L})  \leq  \frac{n!}{n_{\cal S} ! n_{\cal L} ! }   \chi (  q \geq \alpha n_{\cal L} L +n_{\cal S} -1) \left(\sum_{x=1}^{L}  (x/L)^{x-1/2}\right)^{n_{\cal L}}    \left(\sum_{x=1 }^{\lceil\alpha L\rceil}   (x/L)^{x-1/2} \right)^{n_{\cal S}} $$
The sums over the dummy variable $x$ are estimated as
$$
\sum_{x=1 }^{\lceil\alpha L\rceil}      (x/L)^{x-1/2}   \leq CL^{-1/2}, \qquad    \sum_{x=1 }^{L}      (x/L)^{x-1/2}   \leq C 
$$
where the constant $C$ in the first inequality of course depends on $\alpha$ and diverges when $\alpha \to 1$. \qed

%
%
%

Plugging \eqref{eq: conclusion zet} into \eqref{eq: first zet} and using $\sqrt{q} \geq \sqrt{L}$, we get
\be \label{eq: final}
\text{\eqref{eq: m here}}  \, \leq \,   \sum_{n=2 }^{q+1}   \left( \sum_{n_{\cal L} =0}^1  \frac{(C/C_0)^{n}}{n_{\cal S} ! n_{\cal L}!  }  \sum_{L=1}^{q}(L/q)^{q}  +   \sum_{n_{\cal L} =2}^n      \frac{(C/C_0)^{n}}{n_{\cal S} ! n_{\cal L}!  }    \sum_{L=1}^{ \lfloor \frac{q+1}{\alpha n_{\cal L}} \rfloor }     L^{(n_{\cal L}-1)/2} (L/q)^{q}    \right).
\ee
 The first sum over $L$ is trivially bounded by $C$. For the second sum over $L$, we use  $\sum_{L=1}^{M} L^p \leq M^{p+1}$ to get the upper bound
$$
q^{-q}  \left(\frac{q+1}{\alpha n_{\cal L}}\right)^{q+(n_{\cal L}-1)/2+1}   \leq C  (q+1)^{(n_{\cal L}+1)/2}  (\alpha n_{\cal L})^{-(q+1)}  \leq C
$$
provided $\alpha>1/2$ and $1<n_{\cal L}<q+1$.  It follows that \eqref{eq: final} can be made smaller than $1$ by choosing $C_0$ large enough. 

\subsection{Bound on remainder $\delta H'(t)$}
\label{appendix:bound}

We start immediately from the expression 
\eqref{eq:H'} that we repeat here
\be \label{eq:H'repeat}
H'(t)=Q^\dagger H(t) Q -iQ^\dagger \partial_t Q. 
\ee
We will plug in $Q=e^{\Omega}$ with $\Omega=\sum^{n_{\rm max}}_{p=1} \Omega_p$ and estimate the local terms. 
Since there will be terms of any range, it is beneficial to introduce here a weighted norm. If $A=\sum_{I} A_I$ with $A_I$ supported on a finite interval $I \subset \mathbb Z$, then we put, for some $\kappa>0$,
$$
\norm A \norm_{\kappa}  :=\sup_{i}\sum_{I \ni i}  \norm A_I\norm e^{\kappa \str I \str}
$$
where the sup ranges over sites $i$.  Let us estimate the first term of \eqref{eq:H'repeat} in this norm (the second is done in a similar way, as we comment below). 
By expanding the exponential, we have 

\be \label{eq:H'repeat}
Q^\dagger H(t) Q = \sum_{k}  \frac{1}{k!}\sum_{i_1,\ldots, i_k}{\rm{ad}}_{\Omega_{i_k}} \ldots {\rm{ad}}_{\Omega_{i_1}}(H(t)).
\ee
with all $i_p \leq n_{\rm max}$.
An obvious upper bound is 
\be 
\norm Q^\dagger H(t) Q \norm_\kappa \leq  \sum_{k}  \frac{C^k}{k!}\sum_{i_1,\ldots, i_k} \mathcal{R} e^{\kappa \mathcal{R}} W \norm \Omega_{i_k} \norm_l  \ldots \norm \Omega_{i_1} \norm_l \norm H(t)\norm_l
\ee
where 
$\mathcal{R}=R (1+\sum_p i_p)$ bounds the range of the local terms and the combinatorial factor 
$$W=  W(i_1,\ldots, i_k) = 2^kR^{k+1} i_1 (i_1+i_2) \ldots  (i_1+i_2+\ldots+ i_{k} ) 
$$
bounds the number of ways the local terms in the $\Omega_p$ can attach to the operator. Furthermore, we use now the bound  $\norm \Omega_{i_k} \norm_l \leq C e^{-r \str i_k\str}$ (true if $n_{\rm max} \leq n_*$) and we set $M\equiv i_1+i_2+\ldots+ i_{k} $.
Then we use the simple bound 
$$
W(i_1,\ldots, i_k)\leq 2^k R^{k+1}  M^k
$$
and the number of ways to choose $i_1,\ldots, i_k$ subject to given $M$ is bounded by $2^M$. This leads to  
\be \label{eq:H'repeatagain}
\norm Q^\dagger H(t) Q \norm_\kappa \leq  \sum_{k}  \frac{C^kR^{k+1}}{k!}\sum_{M\geq k} \mathcal{R} e^{\kappa \mathcal{R}} M^k 2^M  e^{-r M}h.
\ee
Performing first the sum over $k$ with $M$ fixed  gives
\be \label{eq:H'repeatagain}
\norm Q^\dagger H(t) Q \norm_\kappa \leq   R \sum_{M} \mathcal{R} e^{\kappa \mathcal{R}}e^{M}  (CR)^M  e^{-r M}h
\ee
This is obviously convergent if we choose 
\be \label{eq: constraint r}
r > 1 +\log(CR)+ R\kappa,
\ee
and the bound can be made arbitrarily close to $h$ (the bound on the zero-order term) by increasing $r$. 
To estimate $Q^\dagger \partial_t Q $, we start from the identity
$$
Q^\dagger \partial_t Q= \int_0^1 ds  e^{s\Omega(t)} (\partial_t \Omega) e^{-s\Omega(t)}
$$
 and then we follow the same route as above to bound the integrand for any $0\leq s \leq 1$. Up to prefactors, the result is the same.  Finally, to get to $ \delta H'(t)$, we simply have to omit from \eqref{eq:H'repeat} all terms with 
$$
i_1+i_2+ \ldots +i_k \leq n_{\rm max}.
$$
To find a good estimate on what remains, we choose $r$ a bit larger so that $r-r_0$ is larger than the left hand side of \eqref{eq: constraint r}. Then we can extract a factor $e^{-r_0 i_j}$ so that we get 
$$
\norm \delta H'(t)\norm_{\kappa} \leq C e^{-r_0 n_{\rm max}}.
$$

\end{widetext}

\end{document}